\documentclass{iau}

\usepackage{amsmath}
\usepackage{graphicx}
\usepackage{multirow}
\usepackage{csquotes}

\begin{document}

\lefttitle{I. A. Crawford}
\righttitle{Future Technosignature Searches}

\jnlPage{1}{7}
\jnlDoiYr{2026}
\doival{10.1017/xxxxx}

\aopheadtitle{Proceedings IAU Symposium}
\editors{J. Haqq-Misra \& R. Kopparapu, eds.}

\title{Some Thoughts on the Future of Technosignature Searches: Constraining the Fermi Paradox}

\author{Ian A. Crawford}
\affiliation{Birkbeck College, University of London, UK}

\begin{abstract}
This paper examines how future technosignature searches may constrain competing resolutions of the Fermi paradox, with particular attention to the possibility that technologically capable entities (TCEs) are either intrinsically rare or deliberately concealed. I propose a multi-pronged observational strategy comprising expanded radio and optical SETI, spectroscopic searches for biosignatures and technosignatures in exoplanet atmospheres, astronomical searches for large-scale extraterrestrial engineering, and Solar System searches for extraterrestrial artefacts (SETA). The latter is identified as having a distinctive temporal advantage because it can probe evidence accumulated over Solar System (and perhaps even Galactic) history, rather than requiring temporal overlap with TCEs. In this context, I argue that searches for micron-scale interstellar technological debris (`Arkhipov particles') in lunar and planetary regoliths may provide an additional constraint on past Galactic technological activity. Additionally, I briefly argue that this programme of scientific exploration should be considered alongside its geopolitical and governance implications.
\end{abstract}


\maketitle

\section{Introduction}

There is an apparent paradox at the heart of any contemplation of the prospects for Technologically Capable Entities (TCEs)\footnote{I use this designation, rather than the more common ‘Extraterrestrial Intelligence’ (ETI), to make explicit that it is the results of extraterrestrial technology that we might hope to detect, rather than intelligence {\em per se}. \cite{Kipping2025} have recognized the same issue and propose re-designating the acronym ETI to stand for 'Extraterrestrial Technological Instantiations', but it seems to me that a new acronym may better reinforce the point.  Also, the word ‘entity’ is preferable to ‘life’ given the possibility that non-biological technological entities may in principle exist and could be more common than their putative biological precursors (e.g., \citealt{Wright_etal2022}). Additionally, the common phrase ‘extraterrestrial civilisation’ is also best avoided because the concept of what constitutes a ‘civilisation’ is inherently anthropocentric.}  elsewhere in the universe. This paradox – that we intuitively expect intelligent life to be common but see no evidence for it – is usually named after Enrico Fermi, following his famous question “Where is everybody?” that he asked during a conversation at the Los Alamos National Laboratory in 1950 \citep{Jones1985}. Of course, as it stands, Fermi’s question is just a question rather than a paradox, albeit one that must have an answer. The elevation of the question to a paradox, and its attribution to Fermi, have both been criticised \citep{Gray2015}, and the essence of the question itself long predates Fermi (e.g., \citealt{Moynihan2021}), but as the attribution has stuck, we will use it here.

To my knowledge, the first person to make explicit the paradox lurking behind Fermi’s question was David Viewing \citep{Viewing1975}, and \cite{Brin1983} made the same observation a few years later. In essence, the logic runs as follows: we know that planets are common companions of stars, we know that the laws of physics and chemistry are common throughout the observable universe, we know that these laws permit the origin and evolution of life because life has arisen on Earth, we know that life appeared early in Earth’s history, and we know that the universe is both very large and very old. These considerations would lead us to expect life to be common in the universe. It is true that, on Earth, the evolution of a TCE took about 4 billion years (i.e., most of Earth’s history to-date) which might lead us to suspect that even if life is common, TCEs might nevertheless be rare. But this line of argument runs up against the amount of time available – the average age of rocky planets in the Galaxy is about 7 billion years \citep{Lineweaver2001, Zackrisson_etal2016} so if life is common, and the evolution of technology takes about as long as it did here, then TCEs should still be common, and on average should have a 2.5 billion year head start on us. So where are they? What have they been doing with their (surely by now extremely capable) technologies such that we see no sign of their activities? As Viewing put it: 

\begin{quote}

This then is the paradox: all our logic, all our anti-isocentrism, assures us that we are not unique – that they must be there. And yet we do not see them. \citep{Viewing1975}

\end{quote}

\section{Are TCEs Hiding in Plain Sight?}

Before going any further, we should briefly consider the possibility that evidence of TCEs might be hiding in plain sight, but, perhaps owing to psychological pressures arising within the culture of science, are not recognized as such. This possibility was perhaps articulated most clearly by Stanisław Lem in his science fiction novel {\em Fiasco}:

\begin{quote}

The primary rule of observation said that whatever did not clearly show up as an artificial source had to be considered a natural phenomenon. Astrophysics, besides, had advanced to the point where it possessed sufficient hypotheses to ‘explain’ every kind of observed emission without resorting to the existence of other beings as the senders.

A paradox arose: the greater the number of theories astrophysics had at its disposal, the more difficult it became to prove the authenticity of an intentional signal. \citep[][p.93]{Lem1987}

\end{quote}

I agree that we should be mindful of this possibility, and perhaps revisit some of our current interpretations of astrophysical phenomena with the possibility of artificial explanations in mind. However, I am doubtful that this can be the explanation for the Fermi paradox. It seems more likely that many potential alien activities would be instantly recognized as artificial, and some (e.g., the colonisation of planets within our own solar system in the distant past, including potentially the Earth) would be impossible for us to miss. This seems especially likely when we consider that TCEs will have their own agency, and some subset may positively want to be detected. On the other hand, it is true that if TCEs are deliberately hiding from us they might choose to actively disguise their activities as natural phenomenon. This kind of explanation falls within the ‘zoo hypothesis’ proposed by \cite{Ball1973}, to which we will return below.

There is, of course, a popular argument that so-called Unidentified Anomalous Phenomena (UAP) might represent extraterrestrial TCEs operating on Earth at the present time (for a comprehensive history of this cultural phenomenon, see \citealt{Graff2023}). Clearly, if this is the case then there is no Fermi paradox to resolve – we already have the answer to Fermi’s question but have yet to recognize it as such. However, for reasons discussed elsewhere \citep{Crawford2027}, I don’t think UAP are a plausible solution to the Fermi paradox. There are three main reasons for this: (i) the lack of any logical connection between UAP, whatever they may turn out to be, and extraterrestrial TCEs (I realise that UAP enthusiasts assume this connection a priori, but there is no logical reason to do so); (ii) the lack of any persuasive evidence that UAP represent the activities of TCEs; and (iii) the implication that if UAP are due to the activities of TCEs on Earth today then the Galaxy must be highly populated with TCEs, for which there is no other supporting evidence. Still, although I personally am not convinced that UAP are evidence of extraterrestrial technology, insofar as they represent unexplained phenomena, I agree that they should be studied in the hope that we may learn about new things \cite{Crawford2027}.

\section{Future Observational Approaches}

Clearly, we can only assert that the apparent lack of evidence for TCEs amounts to evidence for their absence if we have sufficiently searched all the relevant parameter space(s). In many respects, it seems clear that we are far from having done so. As discussed in other contributions to this volume, there are multiple (mutually reinforcing) searches that we can, and should, undertake to determine if there really is a paradox to resolve:

•	We should continue and expand radio/optical SETI and extend the search to other multi-messenger domains;

•	Building on the start already made with JWST \citep{Seager_etal2025}, we should undertake increasingly sensitive spectroscopic searches for both biosignatures and technosignatures in exoplanet atmospheres. Future instruments that could support such observations include the Habitable Worlds Observatory (HWO; \cite{Feinberg_etal2024}), proposed optical/IR interferometers such as the Large Interferometer for Exoplanets (LIFE; \cite{Glauser_etal2024}), and, perhaps ultimately, a telescope placed at the gravitational focus of the Sun \citep{Friedman_etal2024}.

•	Building on initial searches (e.g., \citealt{Wright_etal2014, Suazo_etal2022}), we should undertake increasingly sensitive astronomical searches for large-scale extraterrestrial engineering (e.g., Dyson spheres, swarms, etc.).

•	In parallel to these future astronomical observations, we should initiate searches for extraterrestrial artefacts in the Solar System, an undertaking sometimes referred to as the Search for Extraterrestrial Artefacts (SETA).

\section{Advantages of SETA}

To my knowledge, the term SETA was first introduced by \cite{Freitas1983}, although \cite{Foster1972} had earlier made the case for searching for extraterrestrial artefacts in the Solar System. The basic idea also appears much earlier in Arthur C. Clarke’s short story {\em The Sentinel} \citep{Clarke1951} and, more famously, in his novel {\em 2001: A Space Odyssey} \citep{Clarke1968}. As a means of constraining the number of TCEs in the universe, SETA has several advantages over the traditional SETI approach of searching for artificially generated signals in real time. The main limitation of SETI is that it requires the transmitting and receiving entities to (approximately) co-exist in time. Unless TCEs are very long-lived and/or very common, this is unlikely (e.g., \cite{Balbi2021}). And, in any case, numerous, long-lived TCEs are probably incompatible with other aspects of the lack of evidence of on-going extraterrestrial technological activity, what \cite{Brin1983} termed the ‘Great Silence’. Moreover, although some types of large-scale extraterrestrial engineering activities, such as Dyson spheres or swarms (e.g., \citealt{Wright_etal2014, Suazo_etal2022}), may be detectable through astronomical observations, and might be expected to outlast the TCEs that constructed them, it turns out that even these may prove to be short-lived once they fall into disrepair (e.g., \cite{Lacki2025}; see also his contribution elsewhere in this volume).

Although spatially restricted to the confines of the Solar System, SETA has the significant temporal advantage of enabling searches for the activities of TCEs over the 4.5 billion-year history of the Solar System (and, as I will argue below, possibly much longer). These advantages led \cite{Rose2004} to conclude their study of ‘Inscribed matter as an energy-efficient means of communication with an extraterrestrial civilization’ with the observation that 

\begin{quote}
    carefully searching our own planetary backyard may be as likely to reveal evidence of extraterrestrial civilizations as studying distant stars through telescopes. 
\end{quote}

A comprehensive recent discussion of how these searches might be undertaken as we continue to explore the Solar System has been provided by \cite{Haqq-Misra2022}. Although, at first sight, searching for artefacts in the Solar System would seem to only help us determine if the Solar System has been deliberately visited by TCEs, we shouldn’t discount the possibility that this might, at least in principle, have happened multiple times throughout Solar System history.

\section{Some Implications of the Drake Equation}

To put this into perspective, consider an ‘optimistic’ solution to the Drake equation \citep{Drake1965}, where the first 6 terms are all set to unity. Famously, this results in the most optimistic possible solution (at least if we neglect interstellar colonization) such that the number of TCEs extant at the present time, $N_{TCE}$, would equal the average lifetime, $L$, of TCEs (measured in years given that the star formation rate has been assumed to be $\sim 1$ star per year). So, if we adopt $L$ to be 1000 years (possibly optimistic, but who knows?), we would expect $N_{TCE} \sim 1000$ and only about one star in $\sim10^8$ would host a TCE at the present time. The closest would likely be several thousand light-years away, and plausibly not yet detectable by our SETI searches, so the lack of SETI detections could still be consistent with an optimistic solution of the Drake equation.

However, because of the way the Drake equation is constructed, the product of the first six terms equals the formation rate of TCEs, expressed as TCEs per year (see also \citealt{Kipping2025}). So, by assuming an optimistic version of the Drake equation, we have implicitly assumed that one TCE arises every year and, regardless of their average lifetimes, about $10^{10}$ should have appeared over the history of the Galaxy. Clearly, this is an enormous number and would imply that roughly ten percent of all stars has hosted a TCE, even if only an infinitesimal fraction ($\sim10^{-7}$ if $L\sim1000$ years) are extant at the present time. Given that the Solar System is 4.5 billion years old, there would appear to have been ample opportunities for some fraction of the vast number of TCEs that (given ‘optimistic’ assumptions) have existed throughout Galactic history to have visited our Solar System, and for some fraction of those to have left evidence of their visits. This is the main reason for taking SETA seriously — only by searching our own ‘backyard’ can we place constraints on the possibility that spacefaring TCEs have passed this way; if we don’t look, we will never know. 

\section{The Zoo Hypothesis or Nothing?}

Regardless of whether alien artefacts have been left elsewhere in the Solar System, we do at least know, building on the arguments of \cite{Hart1975} and \cite{Viewing1975}, that Earth itself has never been appropriated by TCEs for their own purposes. If this had happened we would not be here now with our own evolutionary history intact. This is the key insight that gives the Fermi paradox its bite — if TCEs have been common in the history of the Galaxy (even if rare at any given moment), isn’t it odd that life on Earth has been allowed to evolve for billions of years without (obvious) interference from outside?

Despite numerous purported solutions to the Fermi paradox over the years (e.g., \citealt{Webb2015, Ćirković2018}), it seems to me that \cite{Ball1973} and \cite{Viewing1975} correctly identified the most plausible scenario able to reconcile a universe where TCEs are (or have been) common with our lack of evidence for them — i.e., that they are deliberately hiding from us, a state of affairs that \cite{Ball1973} dubbed the ‘zoo hypothesis’. This also appears to have been David Brin’s preferred explanation for his ‘Great Silence’ (Brin’s version might perhaps better be dubbed the ‘nursery hypothesis’, but it falls into the same general category of explanations; \cite{Brin1983}). Reluctantly, I agree with Viewing that the only other viable solution for Fermi’s paradox is that TCEs have {\em not} been common in the history of the Galaxy. As \cite{Viewing1975} put it in the abstract to his paper:

\begin{quote}
Possible resolutions [of the paradox] may be either that the occurrence of intelligent life is much rarer than is commonly believed or that our planet occupies a volume of space that has been artificially preserved in a ‘natural’ state.
\end{quote}

Following this logic, \cite{Crawford2024} argued that the solution to Fermi’s question may boil down to ‘the zoo hypothesis or nothing’, where ‘nothing’ would signify that TCEs have been very rare (or, as \cite{Hart1975} argued, non-existent) over the history of the Galaxy. However, the zoo hypothesis also appears incompatible with TCEs being common, because there seems no reason to assume that multiple TCEs would all abide by the same rules for the zoo (this would fall into the 'monocultural fallacy' proposed by \cite{Wright_etal2014}). These considerations imply that ‘optimistic’ solutions to the Drake equation are misplaced — probably because of very low probabilities for one or more of the ‘biological’ factors (e.g., the origin of life, the evolution of ‘complex’ life, the evolution of intelligence, and/or the evolution of a technological culture).

\section{Constraining the Zoo Hypothesis}

If the solution to Fermi’s paradox were to boil down to some version of the zoo hypothesis or to the non-existence of TCEs, it becomes important to ask if the former possibility might produce technosignatures that we could identify. We must assume that sufficiently advanced TCEs would be able to remain hidden if (for whatever reason(s)) they wished to do so. Presumably they could ensure that they do not leave artefacts lying around, and could disguise many of their activities as natural astrophysical processes (recall Stanisław Lem’s argument quoted above). At first sight, therefore, this would seem to leave few options for constraining the zoo hypothesis through technosignature searches.

However, there may be a possible approach building on a suggestion first made by \cite{Arkhipov1996, Arkhipov1998}. Arkhipov’s proposal does not assume that alien artefacts have ever been deliberately directed to the Solar System, but just that any spacefaring TCEs, even if they never leave their own planetary system(s), cannot avoid producing a large amount of space debris. After only ~70 years in space, and very few operations beyond Earth orbit, humanity has already generated copious amounts of debris, and a true space-faring TCE (for example one engaged in mining asteroids for raw materials) would be expected to generate orders of magnitude more. Moreover, clouds of such debris will also be produced by the ‘grinding down’ of abandoned mega-engineering structures \citep{Lacki2025}. The smaller parts of this debris (e.g., micron-sized specks of unusual alloys) will eventually be pushed out into interstellar space by radiation pressure from the parent star. Larger objects, including abandoned spacecraft, may be expelled through gravitational sling shots by giant planets. 

Clearly, the more spacefaring TCEs that have existed over the history of the Galaxy, the greater will be the interstellar density of such debris. Importantly, even if TCEs are maintaining (and perhaps have maintained an astronomically long-lived) quarantine around the Earth, they may have no way of preventing such particles (dubbed `Arkhipov particles' by \cite{Pinault2024}) from drifting into the Solar System from the interstellar medium (ISM). It is therefore conceivable that a careful search for (presumably micron- to sub-micron-sized) fragments of exotic materials in lunar and planetary regoliths could help constrain the number of spacefaring TCEs that have existed in the history of the Galaxy, even if none of them have deliberately visited our Solar System, and even if they have taken steps to hide their activities \citep{Pinault2024, Pinault_etal2026}.

Also difficult to hide would be the rich planetary biospheres from which TCEs (presumably) evolved. It is worth noting that a single planet like the Earth could in principle support multiple (perhaps dozens) of independent biological to technological transitions over a billion or more years of stable habitability after the first appearance of complex life. We may therefore get a hint that we live in a universe of hidden TCEs if rich biospheres turn out to be common, but evidence of intelligence and technology is suspiciously absent. But, on the other hand, such an observation might just mean that one or more great filter(s) exist between complex biology and technology \citep{Hanson1998, Haqq-Misra2020}, and we would be back to the ‘zoo hypothesis or nothing’ ambiguity.

It may be that resolving this ambiguity will eventually require direct exoplanetary investigations with interstellar space probes (e.g., \citealt{Crawford2018, Haqq-Misra2025}). Obviously, this is a task for future generations as far as we are concerned, but raising it brings us forcefully back to the Fermi paradox – Earth has had an observable biosphere for billions of years, but we have no evidence that anyone ever visited us to investigate. And, if they did, they haven’t left any obvious signs of having done so (which brings us back to some version of the zoo hypothesis again).

\section{Some Political Considerations: Who Speaks for Earth?}

There is one further aspect of technosignature deliberations that, to my mind, is often overlooked in scientific and technical discussions of the topic. This is the essentially political question raised by Carl Sagan in the final chapter of {\em Cosmos} \citep{Sagan1980}: {\em Who Speaks for Earth?} 

If SETI detects an alien radio signal, or if alien artefacts are found elsewhere in the Solar System, who can legitimately decide on the next steps? Surely, these decisions, which might conceivably affect the whole future trajectory of life on Earth, cannot be left to individuals, or even individual nation-states, and yet we currently lack the global political architecture required to address them on behalf of humanity as a whole (e.g., \citealt{Michaud2015, Crawford2021}). Currently, some guidelines are provided by the Declaration of Principles Concerning the Conduct of the Search for Extraterrestrial Intelligence drafted by the International Academy of Astronautics \citep{IAA2026}, and there is important on-going community activity in the form of 'post-detection hubs' (e.g., \citealt{Elliot2026}) which may lead to the development of stronger policies. However, all such community-based proposals are entirely voluntary and unenforceable.

The closest we currently have to a global institution that might provide a coordinating role is the United Nations (UN), and it is notable that the IAA SETI Principles require that evidence of extraterrestrial intelligence be reported to the UN Secretary General. However, as currently constituted, the UN has proved to be largely powerless in the face of global crises (e.g., climate change, loss of biodiversity, pandemics, nuclear proliferation, endemic warfare, etc.), mainly owing to the inability of its bickering member nation-states to reach agreement on important global issues. Given this track record, it is hard to have confidence that the UN (again as currently constituted) will be able to provide the political leadership required when, or if, humanity discovers that we are not alone in the universe. To my mind \citep{Crawford2021}, this logic implies the need for much stronger institutions of global governance. 

Of course, it is unrealistic to expect major geopolitical innovations on the off chance that we may one day discover alien technosigntures. However, it is becoming clear that the world will probably need to develop stronger global political institutions (perhaps in the form of a stronger and more democratic UN) to deal more effectively with all the other existential risks humanity will face in the 21st century (e.g., \citealt{Leinen2018, Lopez-Claros2020}). Although seemingly far removed from our main areas of scientific expertise, perhaps the astrobiology community should be prepared to engage with these geopolitical discussions so that we can be ready to contribute when, or if, the time comes. 

\section{Conclusion}

In his famous mid-19th century discourse {\em Of the Plurality of Worlds}, William Whewell observed that:

\begin{quote}
    
The discussions in which we are engaged belong to the very boundary regions of science, to the frontier where knowledge […] ends and ignorance begins. \citep[][p.115]{Whewell1853}

\end{quote}

It is astonishing to reflect that, despite all the progress made in both astronomy and biology since Whewell wrote these words, we are still, at least as far the search for extraterrestrial life is concerned, sitting on the boundary between knowledge and ignorance — we still do not know whether we are alone in the universe or not. 

There is only one way to lessen our ignorance, and that is to explore the universe around us. Thus, we should continue and expand our SETI activities because doing so costs us very little and, even if the probability of success may be low, we can be sure that the probability will be lower if we don’t look. We should expand our efforts to search for life elsewhere in the Solar System, and for biosignatures in exoplanet atmospheres, because this may, eventually, tell us whether non-technological life is common in the universe and whether the origin of life itself is a low probability event. In parallel, we should continue to search for technosignatures in the vicinity of other stars, and for alien artefacts that may have been left in our own Solar System. The search for artificial debris (‘Arkhipov particles’) drifting in from the ISM may be especially worthwhile because of its potential to place limits on the activities of TCEs integrated over Galactic history without assuming that any TCEs deliberately visited the Solar System. Ultimately, we may have to consider building interstellar probes to make in situ observations of any nearby exoplanets for which astronomical evidence for biosignatures and/or technosignatures is ambiguous. As we do so, we should also be mindful of the social and political implications of a discovery of extraterrestrial life (technologically capable or otherwise) and plan for the building of international institutions robust enough to manage such a discovery for the benefit of humanity as a whole.

This is a programme that will occupy us for decades, and perhaps centuries. But only by pursuing such an ambitious programme of exploration will we gradually move the needle from ignorance to knowledge, and perhaps eventually answer Fermi’s insightful but disturbing question.

\end{document}